\documentclass{cjaa}                   % preprint, the final version for publication
\usepackage{graphicx}                   %for PS/EPS graphics inclusion, new
\input{epsf.sty}                        %for PS/EPS graphics inclusion, old
\input{psfig.sty}                       %for PS/EPS graphics inclusion, old

\begin{document}

   \title{Multiwavelength observations of one galaxy in Marano Field
%\footnotetext{$*$ Supported by the National Natural Science Foundation of China.}
}
%   \subtitle{I. Place Your Subtitle Here}

   \volnopage{}      %%preserved for Editor. DOn't remove!
   \setcounter{page}{1}          %%starting page, preserved for Editor. DOn't remove!

   \author{Y. C. Liang
      \inst{1,2}\mailto{}
%% Please move "\mailto{}" to the corresponding author of the paper.
   \and F. Hammer 
      \inst{2}
   \and H. Flores
      \inst{2}
   \and D. Elbaz
       \inst{3}
   \and L. C. Deng
       \inst{1}
   \and F. Ass$\acute{\rm e}$mat   
        \inst{2}   
   \and N. Gruel
        \inst{4}	    
   \and X. Z. Zheng
        \inst{2}   
   \and D. Marcillac    
       \inst{3}
   \and C. J. Cesarsky
       \inst{5}	
      }
   \offprints{Y. C. Liang}                   %% is disabled in fact

   \institute{National Astronomical Observatories, Chinese Academy of Sciences,
             Beijing 100012, China\\
             \email{ycliang@bao.ac.cn}
%% offprint requests will be sent. Note to pair \mailto{} with \email{}
        \and
	GEPI, Observatoire de Paris, Section de Meudon, 92195 Meudon Cedex, France  \\
%             \email{}
        \and
          CEA, Saclay-Service d'Astrophysique, Orme des Merisiers, F91191
             Gif-sur-Yvette, France	    \\
%             \email{}
        \and
	  Department of Astronomy, University of Florida, USA
        \and
          ESO, Karl-Schwarzschild Strase 2, D85748 Garching bei Munchen, Germany
          }

   \date{Received~~; accepted~~}

   \abstract{We report the multiwavelength observations of one intermediate redshift ($z$=0.3884) 
   galaxy in the Marano Field. These data include ISOCAM middle infrared, VLT/FORS2 spectroscopic and photometric data, 
   associated with the ATCA 1.4 GHz radio and ROSAT PSPC X-ray observations from literature. 
   The Spectral Energy Distribution obtained by VLT spectroscopy exhibits its early-type galaxy property, while,
   in the same time, it has obvious [O\,III]5007 emission line.
   The diagnostic diagram from the optical emission line ratios shows its Seyfert galaxy property.
   Its infrared-radio relation follows the correlation of sources detected at
   15 $\mu$m and radio.
   It has a high X-ray luminosity of 1.26$\times$10$^{43}$ ergs s$^{-1}$,
   which is much higher than the general ellipticals with the similar B band luminosity, 
   and is about 2 orders of magnitude higher
   than the derived value from the star forming tracer, the FIR luminosity.
   This means that the X-ray sources of this galaxy are not stellar components, but the AGN is the dominant component.
   }

   \authorrunning{Y. C. Liang et al. }                %author_head in even pages
   \titlerunning{Multiwavelength observations of one galaxy in Marano field }  % title_head in odd pages

   \maketitle

%
%________________________________________________ sections below
%
\section{Introduction}           %% first-level sections will be auto-capitalized
\label{sect:intro}
%\hspace{15pt}%                   %% preserved for Editor

The Marano Field is a $\sim$1 sq. deg. region in the southern sky around $\alpha$(2000)=03$^h$15$^m$09$^s$
and $\delta$(2000)=-55$^{\circ}$13$\arcmin$57$\arcsec$.
A series of papers have presented the multiwavelength data of this field.
The first deep survey in this field was an optical search for quasars, selected by
colour, grism spectra, and optical variability. 64 quasars and AGN were detected at the limiting
magnitude of B$_J$=22 in a survey area of 0.69 deg$^2$ (Marano, Zamorani \& Ziteklli 1988;
Zitelli et al. 1992).  

In the soft X-ray band (0.5-2 keV), the field has been observed for $\sim$56 ksec with the 
Positional Sensitive Proportional Counter (PSPC) on ROSAT.
Zamorani et al. (1999)
presented such X-ray data and the optical identifications (the ESO NTT for a set of U, B, V, R, J, F bands).
In the inner 15$\arcmin$ field, they detected 50 X-ray sources with $S_x>$3.7$\times$10$^{-15}$ ergs cm$^{-2}$ s$^{-1}$.
Among them, 33 are AGNs and two are radio galaxies.

Deep radio surveys at 1.4 GHz and 2.4 GHz were carried out in Marano Field
with the Australia Telescope Compact Array (ATCA). 
At the limiting flux of $\sim$0.2 mJy, 63 sources 
were found at 1.4 GHz and 48 sources at 2.4 GHz (Gruppioni et al. 1997).

However, the overlap of the radio sources with the ROSAT X-ray sources is relatively small. 
A cross-correlation analysis has been presented by Gruppioni et al. (1996).
Three of the five detected X-ray sources have been 
spectroscopically identified: two of them are quasars and the remaining is an early type galaxy. 
This early type galaxy is the object we are studying here. 

XMM-Newton survey was also carried in an 
exposure time $\sim$80 ksec in the Marano Field (Girdke et al. 2003). 
453 new X-ray objects were discovered except the
50 X-ray objects detected by ROSAT.
The optical identifications for XMM-Newton survey have also been done (Lamer et al. 2003).
However, there are no public data available for the individual object in the both observations at this moment.

We have performed the VLT/FORS2 spectroscopic
observation on some objects in the Marano Field. The high quality VLT spectra
must provide important information on the intrinsic properties of the objects.
Also, deep ISOCAM MIR Guaranteed Time Extragalactic Survey has been done in a 90 arcmin$^2$ region. 
The source counts presented by Elbaz et al. (1999) 
have shown strong evolution of the cosmic infrared background. 

In this paper, we study the properties of one 15$\mu$m-selected galaxy, UDSR09, in this field.
We present its high quality VLT/FORS2 optical spectrum and ISOCAM MIR data, associated with 
the X-ray and radio data from literature.
The meaning of UDSR09 marks the ninth slit of the FORS2 spectroscopic observation in the 
Ultra-Deep-Survey-ROSAT field (Liang et al. 2004).
This object is the only one cross-correlated from the multiwavelength data mentioned above. 
We discuss its radio-IR correlation, the X-ray luminosity L$_{\rm X}$ versus B band luminosity L$_{\rm B}$,
L$_{\rm X}$ versus FIR luminosity L$_{\rm FIR}$ as well.
This paper is organized as follows,
The multiwavelength data of the galaxy are presented in Sect.2, 
the results and discussions are given in Sect.3. Sect.4 gives the conclusion.
Throughout this paper, a cosmological model with  $H_0$=70 km s$^{-1}$
Mpc$^{-1}$, $\Omega _M$=0.3 and $\Omega _\Lambda =0.7$ has been adopted.

\section{Multiwavelength data}

\subsection{IR data from ISOCAM}

The very deep ISOCAM follow-up in the Marano field has been done reaching a flux limit of about
50$\mu$Jy. In the 90 arcmin$^2$ field, 75 objects have been detected (Elbaz et
al. 1999; Elbaz et al. 2005, in preparation), 
35 of them have been observed by using VLT/FORS2 spectroscopy (Liang et al. 2004).
Most of the ISOCAM galaxies are star forming galaxies ($>$77\%), rather than AGN 
(Liang et al.2004; Fadda et al. 2002; Elbaz et al. 2002; Franceschini et al. 2003). 
The IR data of the sample galaxy can be found in Table\,1.

\subsection{Optical data from VLT}

Spectrophotometric observations of the ISOCAM-detected galaxies in the Marano Field were obtained
by using the ESO 8m VLT using the FORS2 with R600, I600 at a resolution of
5\AA$~$and covering the possible wavelength range between 5000 and 9000\AA.
The slit width was 1.2{$^{\prime\prime}$} and the slit length was
10$^{\prime\prime}$. Spectra were extracted and wavelength-calibrated using
the IRAF package. Flux calibrations were done using 15 minute
exposures of photometric standard stars (Liang et al. 2004). We study the properties of the galaxy,
UDSR09, the object superimposed on the ninth slit in the VLT observation run.

Using the GIM2D software 
package (Simard et al. 2002), we analyze the optical image of UDSR09 obtained from VLT/FORS2 and 
try to obtain some morphology parameters, including the 
bulge fraction, bulge effective radius, disk scale length and
inclination $i$ (the disk axis to the line of
sight) (see Table\,2).

\subsection{Radio data from ATCA from reference}

Gruppioni et al. (1997) presented the radio observations of the Marano Field
carried with the ATCA at 1.4 and 2.4 GHz with a limiting flux of $\sim$0.2 mJy at each frequency, 
extracted from an area of $\sim$0.36 deg$^2$.
63 objects were detected at 1.4 GHz. One of the galaxies with multiwavelength data available 
is studied in this paper (see Table\,3, Col.(2)-(6)).

\subsection{X-ray data from ROSAT from reference}

The ROSAT PSPC observation in the Marano Field detected
about 50 X-ray sources
in the inner 15 arcmin radius circle at a flux limit of
$\sim$3.7$\times$10$^{-15}$ ergs cm$^{-2}$ s$^{-1}$. 33 of them are AGNs and
two are radio galaxies (Zamorani et al. 1999). 
One of the radio galaxies is the target source in this paper (see Table\,3, Col.(7)-(9)).

%%%%%%%%%%%%%%%%%%%%%%%%%%%%%%%%%%
\section{Results and discussion}
%%%%%%%%%%%%%%%%%%%%%%%%%%%%%%%%%%

The derived results and characteristic parameters for the target galaxy are presented in Table\,1-4.
Col.(1) of all the tables gives the name of the galaxy, UDSR09, following Liang et al. (2004). 

\subsection{IR data}

 Table\,1 illustrates the basic data of the galaxy as follows,
 RA and DEC in 2000 epoch, redshift, ISOCAM 15$\mu$m flux,
 far-infrared luminosity and Star Formation Rate (SFR) derived from IR luminosity.
This galaxy is a luminous infrared galaxy in the $z\sim$0.4 intermediate redshift Universe,
with a high SFR of $\sim$34 $M_\odot$\,yr$^{-1}$.

\subsection{Optical data}

Fig.\,1a shows the optical spectral energy distribution of the galaxy obtained from VLT/FORS2. It shows a elliptical galaxy property,
and has strong metal absorption lines, for example, Ca H K, G-band, Mg H, Na D lines,
but weak Balmer absorption lines. 
In particular, it shows obvious [OIII]5007 emission line, 
as well as [OII]3727 emission line. The H$\beta$ emission line is weak. 

In Table\,2, Col.(2) gives the B band absolute magnitude in the Vega system, Col.(3) gives
the dust extinction by considering energy balance between IR and H$\beta$ luminosities,
Col.(4) and (5) are the emission line ratios, which exhibit its Seyfert 2 galaxy property, which
is shown in the diagnostic diagram of Fig.\,1b (also see Liang et al. 2004). 

We use GIM2D (an exponential disk and a de Vaucouleurs R$^{1\over 4}$ law for bulge) 
to try to obtain the morphology parameters of the galaxy (Col.(6)-(9) in Table\,2).
The interesting thing is the bulge effective radius
R$_e$ is larger than the disk scale length R$_d$, which means the morphological parameters of this galaxy cannot be
decomposed into simple bulge/disk decompositions.

\subsection{Radio-IR correlation}

Table\,4 (Col.(7),(8)) illustrates the 15$\mu$m middle infrared and 1.4 GHz radio luminosities of the galaxy, which
generally follow the radio-IR correlations
presented by Elbaz et al. (2002) (their Fig.9a) and Gruppioni et al. (2003) (their Fig.1) for 15$\mu$m-detected radio galaxies.
Our Fig.\,1c shows such correlation of it.
And the corresponding $q$ value (from Eq.(2) of Elbaz et al. 2002)
is 1.74, a little lower than those of the local galaxies (2.34) and the ISOCAM intermediate-$z$ galaxies (2.3)
shown in Fig.9b of Elbaz et al. (2002).

\subsection{X-ray luminosity versus B band luminosity}

If the X-ray emission sources in the early-type galaxies are stellar in origin, 
the total X-ray luminosity from these sources should scale with B band luminosity L$_{\rm B}$.
While emission from other sources,
such as hot gas, may not be so directly linked to stellar population (Irwain \& Sarazin 1998; O'Sullivan et al. 2001).

Table\,4 gives the derived L$_{\rm X}$ (0.5-2 keV from ROSAT) and L$_{\rm B}$ of the target galaxy.
The galaxy has high L$_{\rm X}$, $\sim$1.26$\times$10$^{43}$ ergs s$^{-1}$, which is much higher
than that of the general early-type galaxies with similar blue band luminosity (see our Fig.\,1d).
We consider that, if all blue stars were embedded in the ionized gas, 
the L$_{\rm B}$ may be underestimated by 3 mag (Liang et al. 2004). This will lead
log\,L$_{\rm B}$ to increase $\sim$1 dex, but still correspond to much lower L$_{\rm X}$ than the observed one at the similar L$_{\rm B}$.

Since the target galaxy shows much higher L$_{\rm X}$ than the general early-type galaxies at the given L$_{\rm B}$,
the source of its X-ray emission may not be the stellar components. 
There must be other sources causing the unusually high X-ray luminosity,
e.g. hot gas or the central buried Active Galactic Nuclei (AGN). 
There may exist a Black Hole at it center.

\subsection{X-ray luminosity versus FIR luminosity: the SFR}

To further understand the sources of X-ray emission, one way is to compare the L$_{\rm FIR}$ with L$_{\rm X}$, hence the derived SFRs.
Ranali et al. (2003) studied the X-ray luninosity as a SFR indicator for the local star forming galaxies:
$
{\rm log}(L_{\rm 0.5-2 keV})=(0.87\pm 0.08){\rm log}(L_{\rm FIR})+2.0(\pm 3.7).
$

The L$_{\rm FIR}$ of the galaxy is $\sim$4.0$\times$10$^{44}$ ergs\,s$^{-1}$ (from L$_{\rm IR}$ by using Eq.(4) of Elbaz et al. 2002).
This corresponds to L$_{\rm X}(\rm 0.5-2 keV)$ $\sim$6.3$\times$10$^{40}$ ergs\,s$^{-1}$ as suggested above.
This is about two orders of magnitude lower than the L$_{\rm X}$ derived directly from 
ROSAT flux (1.26$\times$10$^{43}$ ergs\,s$^{-1}$).
This obvious discrepancy confirms that the X-ray emission of the sample galaxy may not be due to
star forming, there must be other sources responsible for it, e.g. the buried AGN.

{ 
\begin{table*}
\centering
{ \scriptsize
%\tiny
\begin{center}
\caption { The IR data of the sample galaxy}
\end{center}

\label{tab1}

 \begin{tabular}{ccccccc} \hline

Object & RA         & DEC         & z      & f(S$_{15}$)  & log($\frac{L_{\rm IR}}{L_{\odot}}$)  & SFR$_{\rm IR}$   \\
       & (2000)     & (2000)      &        & ($\mu$Jy) &                  &  ($M_{\odot}$\,yr$^{-1}$)      \\ \hline
UDSR09 &  3 14 56.1 & $-55$ 20 08 & 0.3884 & 608.9   &  11.30$\pm$0.08  &  34.21                        \\ \hline

 \end{tabular}
}
\end{table*}
}

{ 
\begin{table*}
{ \scriptsize
%\tiny
\caption { The optical data: emission line ratios and morphology parameters}

\label{tab2}

 \begin{tabular}{ccccc|cccc} \hline

Object  &  M$_{\rm B}$ & A$_V$  & log($\frac{\ion{[O}{ii]}}{H\beta}$ & log($\frac{\ion{[O}{iii]}}{H\beta}$) & b   & Re   & Rd   & $i$  \\
  (1)   &   (2)        &   (3)  &     (4)                            & (5)                                  & (6) & (7)  & (8)  &  (9)  \\ \hline
 UDSR09 &  -21.58      & 3.99   & 1.61$\pm$0.09                      & 1.35$\pm$0.03    &0.77$\pm$0.02 &  5.61$\pm$0.30 & 3.34$\pm$0.30 & 57$\pm$5    \\ \hline
 \end{tabular}
} \\
{\tiny Notes: b refers bulge fraction; Re, bulge effective radius; Rd, exponential disk scale length; 
$i$, disk inclination.}
\end{table*}
}

{ 
\begin{table*}
{ \scriptsize
%\tiny
\caption { The radio and X-ray data from literature}

\label{tab3}

 \begin{tabular}{c|ccccc|ccc} \hline
       &        &                     &   Radio$^a$              &                    &                      &        &     X-ray$^b$       &         \\ \hline
Object & object & S$_{\rm Peak,1.4}$  & S$_{\rm Total,1.4}$  & S$_{\rm Peak,2.4}$ & S$_{\rm Total,2.4}$  & object & Net counts & S$_x$    \\
       &        & (mJy)               &  (mJy)               &     (mJy)          &  (mJy)               &                     &            &
       (10$^{-14}$ergs~cm$^{-2}$~s$^{-1}$)       \\
 (1)   &    (2) &    (3)              &    (4)               &  (5)               &      (6)             &     (7)             &  (8)       &  (9)     \\ \hline
UDSR09 &  38B   & 0.36$\pm$0.03       & 0.57$\pm$0.08        & 0.43$\pm$0.04      & 0.44$\pm$0.12        & X021-5 & 102.3      & 2.36$\pm$0.25        \\ \hline

 \end{tabular}
} 
{ \scriptsize Notes: ``$a$" is for (Gruppioni et al. 1997) and ``$b$" is for ROSAT PSPC from Zamorani et al. (1999). 
 Col.(2) and (7) are the corresponding names in the two references.} .
\end{table*}
}

{ 
\begin{table*}
{ \scriptsize
%\tiny
\caption { Some derived luminosities and parameters}

\label{tab3}

 \begin{tabular}{ccccccccc} \hline

Object & log($\frac{L_{\rm B}}{L_{\odot}}$)           & flux(\ion{[O}{iii]})      &  L$_{\ion{[O}{iii]}}$  & L$_{\rm X}$      &  L$_{\rm IR}$    &  $\nu$L$_{\nu}$[15$\mu$m] & $\nu$L$_{\nu}$[1.4GHz] & q    \\
       &                 & (ergs~cm$^{-2}$~s$^{-1}$) &  (ergs~s$^{-1}$)       & (ergs~s$^{-1}$)  & (ergs~s$^{-1}$)  &  (L$_{\odot}$)             & (L$_{\odot}$)          &      \\ 
 (1)   &    (2)          &    (3)                    &    (4)                 &  (5)             &      (6)         &     (7)                    &  (8)                   &  (9)     \\ \hline
UDSR09 &    9.98        & 9.34$\times$10$^{-17}$   &  4.9$\times$10$^{40}$  & 1.26$\times$10$^{43}$ & 7.63$\times$10$^{44}$ & 2.13$\times$10$^{10}$    & 6.95$\times$10$^{5}$  &  1.74       \\ \hline

 \end{tabular}
} 

\end{table*}
}

\begin{figure*}
% \centering
   \includegraphics[bb=55 317 436 669,width=14.6cm,clip]{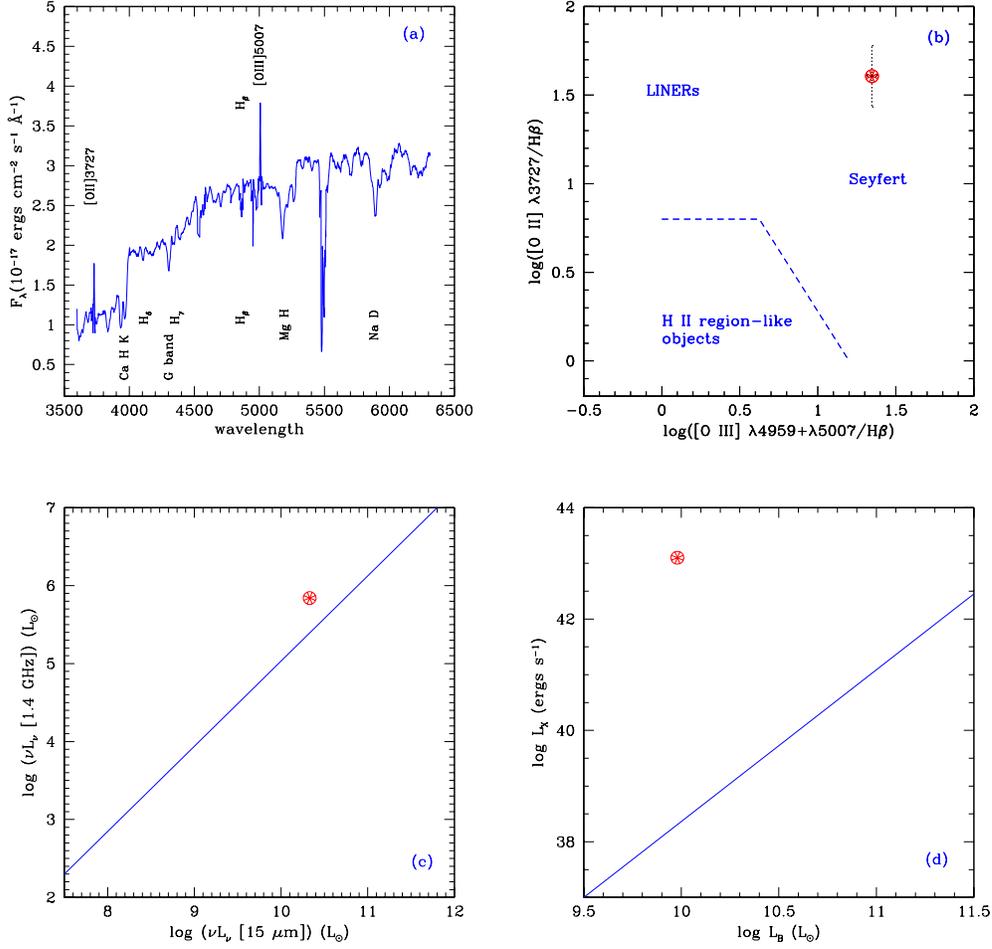}

\caption {{\bf (a)} The VLT/FORS2 optical spectrum of the galaxy (the top left); 
{\bf (b)} The diagnostic diagram from optical emission line ratios (the top right); 
{\bf (c)}  The radio-IR correlation (the bottom left), and the solid line is from Gruppioni et al. (2001)
for the best-fitting obtained for the radio sources detected by 15$\mu$m in the ELATS southern fields;
{\bf (d)} The X-ray luminosity (0.5-2 keV) versus the blue optical luminosity (the bottom right), and the solid line
is taken from Fig.1 of Irwin \& Sarazin (1998) for the best-fitting power-law relation for early-type galaxies.
 }
\label{LXLBs}

\end{figure*}

%
%      set size-flexible figure with epsf.sty: another old way
%------------------------------------------------------------ Fig2: S/N curve

\section{Conclusions}
\label{sect:conclusion}
%\hspace{15pt}%                   %% preserved for Editor

We present the multiwavelength data of one galaxy in Marano Field, including the ISOCAM middle infrared, VLT spectroscopic and photometric data obtained
by us, associated with the ATAC radio 1.4 GHz and ROSAT PSPC X-ray data (0.5-2 keV) from the literature. 

The high quality VLT spectrum show the obvious [OIII] emission line.
The [OIII], [OII] and H${\beta}$ emission lines diagnostic diagram shows its Seyfert 2 galaxy property.
Its IR and radio data relation follows the radio-IR correlations of local and ISOCAM intermediate-$z$ galaxies.

However, its X-ray luminosity is much higher than that of general early-type galaxies at the given B band luminosity, 
and much higher than the star forming tracer, FIR luminositys,
which means that the sources of its X-ray emission
should not come from stellar components, and the AGN is the dominate component.

\begin{acknowledgements}
We thank Dr. Stefanie Komossa for her very valuable suggestions, which helped to improve this work. 
We thank Mr. James Wicker for his help to improve the English language in the text.
This work was funded by the Natural Science Foundation of China (NSFC) under 
Nos.10403006 and 10433010.
\end{acknowledgements}

%\appendix                  %%appendicial material is supported

\label{lastpage}


\begin{thebibliography}{99}
%% you can type \apj for ApJ, \aap for A&A, \apss for Ap&SS, etc. Please consult
%% the macro cjaa.cls. You can also find them in aasguide.tex (AASTeX for ApJ, AJ, PASP)
%% Please follow the format of ChJAA's reference list

\bibitem[2004] {Elbaz04} Elbaz D. et al. 2005,
 in preparation

\bibitem[2002] {Elbaz02} Elbaz D., Cesarsky C. J., Chanial P. et al. 2002,
\aap, 384, 848

\bibitem[2002] {Fadda02} Fadda D., Flores H., Hasinger G. et al. 2002,
\aap, 383, 838

\bibitem[2003] {F03} Franceschini A., Berta S., Rigopoulou D. et al. 2003,
\aap, 403, 501


\bibitem[1999] {G03} Giedke K., Wilms J., Lamer G. et al. 2003,
Astron. Nachr./AN 324, 136

\bibitem[1999] {G96} Gruppioni C., Zamorani G., de Ruiter, H. R. et al. 1996,
MPE Report, 263, 653 (1996rftu.proc., 653)

\bibitem[1997] {G97} Gruppioni C., Zamorani G., de Ruiter H. R. et al. 1997,
MNRAS, 286, 470

\bibitem[1999] {G03} Gruppioni C., Pozz, F., Zamorani G., 2003,
MNRAS, 341, L1

\bibitem[1998] {I98} Irwin J. A. \& Sarazin C. L. 1998,
\apj, 499, 650

\bibitem[1999] {G03} Lamer G., Wagner S., Zamorani, G. et al. 2003,
Astron. Nachr./AN 324, 16


\bibitem[1996]  {Lemke96} Lemke D., Klaas U., Abolins J. et al. 1996,
\aap, 315, L64


%\bibitem[2004] {liang2004} Liang Y. C., Hammer F., Flores H. et al. 2004, 
%\aap, 417, 905

\bibitem[2004] {liang2004} Liang Y. C., Hammer F., Flores H. et al. 2004, 
\aap, 423, 867

\bibitem[1988] {Marano88} Marano B, Zamorani G., Ziteklli V., 1988,
MNRAS, 232, 111


\bibitem[1994] {M94} Mulchaey, J. S., Koratkar, A., Ward M. J. et al. 1994,
\apj, 436, 586

\bibitem[2001] {OS01} O'Sullivan E., Forbes D. A., Ponman T. J. 2001,
MNRAS, 328, 461

\bibitem[2003] {R03} Ranalli P., Comastri A. and Setti G. 2003,
\aap, 399, 39
 

\bibitem[2002] {simard02} Simard L., Willmer C. N. A., Vogt N. P. et al. 
2002, \aaps 142, 1


\bibitem[1999] {Z99} Zamorani G., Mignoli M., Hasinger G. et al. 1999,
\aap, 346, 731


\bibitem[1999] {Z92} Zitelli V., Mignoli M., Zamorani G. et al. 1992,
MNRAS, 256, 349
  

\end{thebibliography}
\end{document}